# Geometric algebra, qubits, geometric evolution, and all that

*Alexander M. SOIGUINE*

**Abstract:** The approach initialized in [1], [2] is used for description and analysis of qubits, geometric phase parameters – things critical in the area of topological quantum computing [3], [4]. The used tool, Geometric (Clifford) Algebra [5], [6], is the most convenient formalism for that case. Generalizations of formal "complex plane" to an arbitrary variable plane in 3D, and of usual Hopf fibration to the map generated by an arbitrary unit value element $g_0 \in G_3^+$ are resulting in more profound description of qubits compared to quantum mechanical Hilbert space formalism.

## 1. Introduction

The working environment will be even subalgebra $G_3^+$ of elements $\alpha + I_S \beta$ [1], $\alpha$ and $\beta$ are (real[2]) scalars, $I_S$ is a unit size oriented area, bivector, in an arbitrary given plane $S \subset E_3$. It was explained in [7], [8] that elements $\alpha + I_S \beta$ only differ from what is traditionally called "complex numbers" by the fact that $S \subset E_3$ is an arbitrary plane. Traditional "imaginary unit" $i$ is geometrically some $I_S$ when it is not necessary to specify the containing plane – everything is going on in one fixed plane, not in 3D world.

## 2. Rotations with elements of $G_3^+$ and generalized Hopf fibration

Let's take a unit value element of $G_3^+$: $so(\alpha, \beta, S) \equiv$
$\alpha + \beta(b_1 e_2 e_3 + b_2 e_3 e_1 + b_3 e_1 e_2) = \alpha + \beta_1 e_2 e_3 + \beta_2 e_3 e_1 + \beta_3 e_1 e_2, \beta_i = \beta b_i, \alpha^2 + \beta^2 = 1$ and $b_1^2 + b_2^2 + b_3^2 = 1$. If $C$ is a bivector, $C = C_1 e_2 e_3 + C_2 e_3 e_1 + C_3 e_1 e_2$, then the map $so(\alpha, \beta, S)^{\sim} C so(\alpha, \beta, S) \to C_{rot}$, $so(\alpha, \beta, S)^{\sim} = \alpha - I_S \beta$, is rotation of $C$. This rotation generates a map $S^3 \to S^2 : so(\alpha, \beta, S) \xrightarrow{C} C_{rot}$. If, for example, $C = e_2 e_3$ then:

$$(\alpha - I_S \beta) e_2 e_3 (\alpha + I_S \beta) = (\alpha - \beta(b_1 e_2 e_3 + b_2 e_3 e_1 + b_3 e_1 e_2)) e_2 e_3 (\alpha + \beta(b_1 e_2 e_3 + b_2 e_3 e_1 + b_3 e_1 e_2)) =$$

$$(\alpha - \beta_1 e_2 e_3 - \beta_2 e_3 e_1 - \beta_3 e_1 e_2) e_2 e_3 (\alpha + \beta_1 e_2 e_3 + \beta_2 e_3 e_1 + \beta_3 e_1 e_2) =$$

$$(\alpha^2 + \beta_1^2 - \beta_2^2 - \beta_3^2) e_2 e_3 + 2(\alpha \beta_3 + \beta_1 \beta_2) e_3 e_1 + 2(\beta_1 \beta_3 - \alpha \beta_2) e_1 e_2, \quad (2.1)$$

which is classical Hopf fibration $S^3 \to S^2$.

---

[1] This sum bears the sense *"something <u>and</u> something"*. It is not a sum of similar elements giving another element of the same kind, see [1], [9].
[2] Scalars should always be real. "Complex" scalars are not scalars.

Also (2.1) shows that "imaginary" number $i$ in this Hopf fibration written in traditional terms should geometrically be unit value oriented area $e_2 e_3$ orthogonal to the basis vector $e_1$.

Suppose bivector is expanded not in $\{e_2e_3, e_3e_1, e_1e_2\}$ but in any basis of bivectors $\{B_1, B_2, B_3\}$, where $B_1$ is arbitrary unit bivector in 3D, and $B_2$, $B_3$ are two unit bivectors orthogonal to $B_1$ and to each other. Their mutual orientation should satisfy the same multiplication rules as $\{e_2e_3, e_3e_1, e_1e_2\}$ do: $B_1 B_2 = -B_3, B_1 B_3 = B_2, B_2 B_3 = -B_1$. The Hopf fibration then can be extended to an arbitrary generating element from $G_3^+$:

$$so(\alpha, \beta, S) \xrightarrow{C_0 + C_1 B_1 + C_2 B_2 + C_3 B_3}$$

$$C_0 + \left(C_1[(\alpha^2 + \beta_1^2) - (\beta_2^2 + \beta_3^2)] + 2C_2(\beta_1\beta_2 - \alpha\beta_3) + 2C_3(\alpha\beta_2 + \beta_1\beta_3)\right)B_1 +$$

$$\left(2C_1(\alpha\beta_3 + \beta_1\beta_2) + C_2[(\alpha^2 + \beta_2^2) - (\beta_1^2 + \beta_3^2)] + 2C_3(\beta_2\beta_3 - \alpha\beta_1)\right)B_2 +$$

$$\left(2C_1(\beta_1\beta_3 - \alpha\beta_2) + 2C_2(\alpha\beta_1 + \beta_2\beta_3) + C_3[(\alpha^2 + \beta_3^2) - (\beta_1^2 + \beta_2^2)]\right)B_3 \qquad (2.2)$$

3. Quantum mechanical qubits and elements of $G_3^+$

A pure qubit state in terms of conventional quantum mechanics is two dimensional unit value vector with "complex" value components:

$$|\psi\rangle = \begin{pmatrix} z_1 \\ z_2 \end{pmatrix} = z_1 \begin{pmatrix} 1 \\ 0 \end{pmatrix} + z_2 \begin{pmatrix} 0 \\ 1 \end{pmatrix}, z_1^2 + z_2^2 = z_1 \tilde{z}_1 + z_2 \tilde{z}_2 = 1, z_k = z_k^1 + i z_k^2, k = 1,2$$

Let's find explicit relations between elements $|\psi\rangle$ as two-component "complex" vectors and elements $so(\alpha, \beta, S)$ belonging to $G_3^+$. Take an arbitrary $so(\alpha, \beta, S)$. If bivector $e_2 e_3$ is chosen as the one defining "complex plane" $S$, we have:

$$so(\alpha, \beta, S) = (\alpha + \beta_1 e_2 e_3) + \beta_2 e_2 e_3 e_1 e_2 + \beta_3 e_1 e_2 = (\alpha + \beta_1 e_2 e_3) + (\beta_3 + \beta_2 e_2 e_3) e_1 e_2 \equiv z_1^{2,3} + z_2^{2,3} e_1 e_2$$

So we get correspondence:

$$so(\alpha, \beta, S) \to |\psi\rangle = \begin{pmatrix} z_1^{2,3} \\ z_2^{2,3} \end{pmatrix} = \begin{pmatrix} \alpha + i\beta_1 \\ \beta_3 + i\beta_2 \end{pmatrix}, i = e_2 e_3$$

Actually, the number of such maps is infinite because "complex" plane can be an arbitrary plane in 3D. Take arbitrary mutually orthogonal unit bivectors $\{B_1, B_2, B_3\}$ satisfying multiplication rules as above. If $S^3 \ni so(\alpha, \beta, S) \equiv \alpha + I_S \beta$ is expanded in basis $\{B_1, B_2, B_3\}$:

$$\alpha + I_S \beta = \alpha + \beta(b_1 B_1 + b_2 B_2 + b_3 B_3) = \alpha + \beta_1 B_1 + \beta_2 B_2 + \beta_3 B_3, \beta_i = \beta b_i,$$

then, for example, taking $B_1$ as "complex" plane we get:

$$\alpha + \beta_1 B_1 + \beta_2 B_2 + \beta_3 B_3 = \alpha + \beta_1 B_1 + \beta_2 B_1 B_3 + \beta_3 B_3 = \alpha + \beta_1 B_1 + (\beta_3 + \beta_2 B_1) B_3,$$

and the correspondence is:

$$so(\alpha, \beta, S) \to |\psi\rangle = \begin{pmatrix} z_1^{B_1} \\ z_2^{B_1} \end{pmatrix} = \begin{pmatrix} \alpha + i\beta_1 \\ \beta_3 + i\beta_2 \end{pmatrix}, i = B_1 \text{ - arbitrary bivector in 3D} \quad (3.1)$$

So, for any $so(\alpha, \beta, S)$ we have the following mappings:

$$so(\alpha, \beta, S) = \alpha + \beta(b_1 B_1 + b_2 B_2 + b_3 B_3) = \alpha + \beta_1 B_1 + \beta_2 B_2 + \beta_3 B_3 =$$
$$\alpha + \beta_1 B_1 + (\beta_3 + \beta_2 B_1) B_3 =$$
$$\alpha + \beta_2 B_2 + (\beta_1 + \beta_3 B_2) B_1 =$$
$$\alpha + \beta_3 B_3 + (\beta_2 + \beta_1 B_3) B_2$$

$$\Rightarrow \begin{cases} \begin{pmatrix} \alpha + i\beta_1 \\ \beta_3 + i\beta_2 \end{pmatrix}, i = B_1; B_3 \text{ forgotten} \\ \begin{pmatrix} \alpha + i\beta_2 \\ \beta_1 + i\beta_3 \end{pmatrix}, i = B_2; B_1 \text{ forgotten} \\ \begin{pmatrix} \alpha + i\beta_3 \\ \beta_2 + i\beta_1 \end{pmatrix}, i = B_3; B_2 \text{ forgotten} \end{cases} \quad (3.2)$$

All that means that to recover, or establish, which $so(\alpha, \beta, S)$ in 3D is associated with $|\psi\rangle = \begin{pmatrix} z_1 \\ z_2 \end{pmatrix}$ it is necessary, firstly, to define which plane in 3D should be taken as "complex" plane and then to choose another plane, orthogonal to the first one. The third orthogonal plane is then defined by the first two up to orientation that formally means $B_1 B_2 B_3 = \pm 1$.

## 4. New definitions

I shall give new definitions of states, observables, measurements and observable values in the case of 3D objects identifiable as $G_3^+$ elements:

*Definition 4.1.*

> **States** of a qubit and **observables** of a qubit are unit value elements of $G_3^+$. The notations will be used in the following mainly as:

**State** - $so(\alpha, \beta, S) \equiv \alpha + I_S \beta = \alpha + \beta(b_1 B_1 + b_2 B_2 + b_3 B_3) = \alpha + \beta_1 B_1 + \beta_2 B_2 + \beta_3 B_3, \beta_i = \beta b_i$, $\alpha^2 + \beta^2 = 1$, $b_1^2 + b_2^2 + b_3^2 = 1$

**Observable** - $C = C_0 + C_1 B_1 + C_2 B_2 + C_3 B_3, C_0^2 + C_1^2 + C_2^2 + C_3^2 = 1$

*Definition 4.2.*

**Measurement of observable $C$, measured in state $\alpha + I_S \beta$, is a generalized Hopf fibration generated by $C$:**

$$G_3^+ \xrightarrow{G_3^+} G_3^+ : \alpha + I_S \beta \xrightarrow{C} (\alpha - I_S \beta) C (\alpha + I_S \beta),$$

**which is a new element from $G_3^+$ explicitly given in basis components by (2.2).**

*Definition 4.3.*

**Value of observable $C$, measured in state $\alpha + I_S \beta$, is a map of the measurement of the observable to a set of measurement values.**

## 5. Qubit states in $G_3^+$ corresponding to quantum mechanical basis states

Quantum mechanical pure qubit state is $|\psi\rangle = z_1|0\rangle + z_2|1\rangle$, linear combination of two basis states $|0\rangle$ and $|1\rangle$. In $G_3^+$ terms these two states are, as follows from (3.1), the classes of equivalence:

- State $|0\rangle$ corresponds to any one of the following sets of $so(\alpha, \beta, S)_{|0\rangle}$ elements:
    - $\alpha + i\beta_1$, $\alpha^2 + \beta_1^2 = 1$, if the "complex plane" is selected as $i = B_1$, or:
    - $\alpha + i\beta_2$, $\alpha^2 + \beta_2^2 = 1$, if the "complex plane" is selected as $i = B_2$, or: (5.1)
    - $\alpha + i\beta_3$, $\alpha^2 + \beta_3^2 = 1$, if the "complex plane" is selected as $i = B_3$.
- State $|1\rangle$ corresponds to any one the following sets of $so(\alpha, \beta, S)_{|1\rangle}$ elements:
    - $(\beta_3 + i\beta_2) B_3$, $\beta_3^2 + \beta_2^2 = 1$, if the "complex plane" is selected as $i = B_1$, or:
    - $(\beta_1 + i\beta_3) B_1$, $\beta_1^2 + \beta_3^2 = 1$, if the "complex plane" is selected as $i = B_2$, or: (5.2)
    - $(\beta_2 + i\beta_1) B_2$, $\beta_2^2 + \beta_1^2 = 1$, if the "complex plane" is selected as $i = B_3$.

For any of the states $\alpha + \beta_i B_i, i = 1,2,3$, corresponding to $|0\rangle$, the value of observable $B_i$ is:

$$(\alpha - \beta_i B_i) B_i (\alpha + \beta_i B_i) = (\alpha^2 + \beta_i^2) B_i = B_i \qquad (5.3)$$

For any of the states $\beta_{(i+2)\mod 3}B_{(i+2)\mod 3} + \beta_{(i+1)\mod 3}B_{(i+1)\mod 3}, i=1,2,3$, corresponding to $|1\rangle$ (with the agreement 3mod3=3, since index 0 does not here exist), the value of observable $B_i$ is:

$$(-\beta_{(i+2)\mod 3}B_{(i+2)\mod 3} - \beta_{(i+1)\mod 3}B_{(i+1)\mod 3})B_i(\beta_{(i+2)\mod 3}B_{(i+2)\mod 3} + \beta_{(i+1)\mod 3}B_{(i+1)\mod 3}) = $$
$$-(\beta^2_{(i+2)\mod 3} + \beta^2_{(i+1)\mod 3})B_i = -B_i \quad (5.4)$$

This is the actual meaning of quantum mechanical basis states:

**Value of observable $B_i$ is, for any pure qubit state from the set of all $so(\alpha, \beta, S)_{|0\rangle}$, the bivector $B_i$ itself:**

$$so(\alpha, \beta, S)_{|0\rangle} \xrightarrow{B_i} B_i$$

**Value of observable $B_i$ is, for any pure qubit state from the set of all $so(\alpha, \beta, S)_{|1\rangle}$, flipped bivector $-B_i$:**

$$so(\alpha, \beta, S)_{|1\rangle} \xrightarrow{B_i} -B_i$$

Lets' take arbitrary bivector observable $C = C_1B_1 + C_2B_2 + C_3B_3$. Without losing of generality we can think that "complex" plane is defined by bivector $B_1$. Then generalized Hopf fibrations, measurements of the observable $C$, for the states $so(\alpha, \beta, S)_{|0\rangle}$ and $so(\alpha, \beta, S)_{|1\rangle}$, are correspondingly:

$$(\alpha - \beta_1 B_1)C(\alpha + \beta_1 B_1) = C_1B_1 + [C_2(\alpha^2 - \beta_1^2) - 2C_3\alpha\beta_1]B_2 + [C_3(\alpha^2 - \beta_1^2) + 2C_2\alpha\beta_1]B_3 =$$
$$C_1B_1 + (C_2\cos 2\varphi - C_3\sin 2\varphi)B_2 + (C_2\sin 2\varphi + C_3\cos 2\varphi)B_3 \text{ (through parameterization } \alpha = \cos\varphi,$$
$$\beta_1 = \sin\varphi)$$

and:

$$(-\beta_2 B_2 - \beta_3 B_3)C(\beta_2 B_2 + \beta_3 B_3) = -C_1B_1 + [C_2(\beta_2^2 - \beta_3^2) + 2C_3\beta_2\beta_3]B_2 + [2C_2\beta_2\beta_3 - C_3(\beta_2^2 - \beta_3^2)]B_3 =$$
$$-C_1B_1 + (C_2\cos 2\vartheta + C_3\sin 2\vartheta)B_2 + (C_2\sin 2\vartheta - C_3\cos 2\vartheta)B_3 \text{ (through parameterization}$$
$$\beta_2 = \cos\vartheta, \beta_3 = \sin\vartheta).$$

We get the following:

**Measurement of observable $C = C_1B_1 + C_2B_2 + C_3B_3$ in pure qubit state $so(\alpha, \beta, S)_{|0\rangle}$, is bivector with the $B_1$ component equal to unchanged value $C_1$. The $B_2$ and $B_3$ measurement components are equal to $B_2$ and $B_3$ components of $C$ rotated by angle $2\varphi$ defined by $\alpha = \cos\varphi$ and $\beta_1 = \sin\varphi$, where plane of rotation is $B_1$.**

**Measurement of observable** $C = C_1 B_1 + C_2 B_2 + C_3 B_3$ **in pure qubit state** $so(\alpha, \beta, S)_{|1\rangle}$**, is bivector with the** $B_1$ **component equal to flipped value** $-C_1$ **(qubit flips in** $B_1$ **plane). The** $B_2$ **and** $B_3$ **measurement components are equal to** $B_2$ **and** $B_3$ **components of** $C$ **rotated by angle** $2\vartheta$ **defined by** $\beta_2 = \cos\vartheta$, $\beta_3 = \sin\vartheta$**, where plane of rotation is** $B_1$**. The absolute value of angle of rotation is the same as for** $so(\alpha, \beta, S)_{|0\rangle}$ **but the rotation direction is opposite to the case of** $so(\alpha, \beta, S)_{|0\rangle}$.

The above results are geometrically pretty clear. The two states, $so(\alpha, \beta, S)_{|0\rangle}$ and $so(\alpha, \beta, S)_{|1\rangle}$ represented in $G_3^+$ correspondingly by $\alpha + B_1\beta_1$ and $(\beta_3 + B_1\beta_2)B_3$, only differ by additional factor $B_3$ in $so(\alpha, \beta, S)_{|1\rangle}$. That means that measurements of bivector observable $C$ in states $so(\alpha, \beta, S)_{|0\rangle}$ and $so(\alpha, \beta, S)_{|1\rangle}$ are equivalent up to additional "wrapper" $B_3$:

$$so(\alpha, \beta, S)\tilde{}_{|1\rangle} C so(\alpha, \beta, S)_{|1\rangle} = \tilde{B}_3 so(\alpha, \beta, S)\tilde{}_{|0\rangle} C so(\alpha, \beta, S)_{|0\rangle} B_3 \quad (5.5)$$

That simply means that the measurement on the left side is received from the $so(\alpha, \beta, S)\tilde{}_{|0\rangle} C so(\alpha, \beta, S)_{|0\rangle}$ measurement just by flipping of the latter relative to the plane $B_3$.

## 6. Berry parameters for the $G_3^+$ qubit states

Quantum mechanical transformations $|\psi\rangle \to e^{i\theta}|\psi\rangle$, considered as transformations on $S^3$, are often called Clifford translations. Clifford translation of state $so(\alpha, \beta, S)$ with explicitly defined "complex" plane $Cl$ should be written as $so(\alpha, \beta, S) \to e^{I_{Cl}\theta} so(\alpha, \beta, S)$.

Let's take transformation:

$$so(\alpha, \beta, S) \to e^{-Ht} so(\alpha, \beta, S) = e^{-\frac{H}{|H|}(|H|)t} so(\alpha, \beta, S) \equiv so(\alpha, \beta, S, t) \quad (6.1)$$

where $H$ is constant value generic Hamiltonian of the system, bivector of $G_3^+$ with the plane not coinciding with $S$. Clearly, $so(\alpha, \beta, S, t)$ satisfies Schrödinger equation:

$$\frac{d}{dt} so(\alpha, \beta, S, t) = -\frac{H}{|H|}\left(|H| so(\alpha, \beta, S, t)\right)$$

with unit bivector $\frac{H}{|H|}$ in 3D replacing "imaginary unit" $i$ of the conventional quantum mechanics case.

**This is important thing: in terms of** $G_3^+$ **the Schrödinger equation is equation for the result of**

**transformation of a $G_3^+$ qubit state $so(\alpha, \beta, S) \to e^{-Ht} so(\alpha, \beta, S)$ generated by $G_3^+$ bivector Hamiltonian which is generic Hamiltonian of the system.**

Let's consider the Hamiltonian corresponding to constant value $|H|$ magnetic field slowly rotating around axis orthogonal, for example, to plane $B_3$ (axis along vector $e_3$). Initial plane of the field $H$ is inclined by some angle $\theta$ relative to vector $e_3$. It may be initialized as rotation of $B_3$ in plane $B_1$:

$$|H|B_3 \to e^{-B_1\frac{\theta}{2}} |H| B_3 e^{B_1\frac{\theta}{2}}$$

The inclined field rotates around $e_3$ depending on angle $\varphi(t)$:

$$H(t) = |H| e^{-B_3 \frac{\varphi(t)}{2}} e^{-B_1 \frac{\theta}{2}} B_3 e^{B_1 \frac{\theta}{2}} e^{B_3 \frac{\varphi(t)}{2}} \tag{6.2}$$

Calculate (6.2) in two steps. First, incline $B_3$: $|H|B_3 \to e^{-B_1\frac{\theta}{2}} |H| B_3 e^{B_1\frac{\theta}{2}}$ that gives $|H|(-B_2 \sin\theta + B_3 \cos\theta)$. Then rotate the inclined bivector:

$$H(t) = |H| e^{-B_3 \frac{\varphi(t)}{2}} e^{-B_1 \frac{\theta}{2}} B_3 e^{B_1 \frac{\theta}{2}} e^{B_3 \frac{\varphi(t)}{2}} = |H| e^{-B_3 \frac{\varphi(t)}{2}} (-B_2 \sin\theta + B_3 \cos\theta) e^{B_3 \frac{\varphi(t)}{2}} =$$

$$|H|[B_1 \sin\theta \sin\varphi(t) - B_2 \sin\theta \cos\varphi(t) + B_3 \cos\theta] \tag{6.3}$$

Let's now calculate Berry curvature and connection in full $G_3^+$ terms. Forgetting about $|H|$ and using notation $U$ for $|H|[B_1 \sin\theta \sin\varphi(t) - B_2 \sin\theta \cos\varphi(t) + B_3 \cos\theta]$ we have:

$$\tilde{U} = -B_1 \sin\theta \sin\varphi + B_2 \sin\theta \cos\varphi - B_3 \cos\theta = e^{-B_3 \varphi} B_2 \sin\theta - B_3 \cos\theta,^3$$

$$\frac{\partial U}{\partial \theta} = -(e^{-B_3 \varphi} B_2 \cos\theta + B_3 \sin\theta), \quad \text{then} \quad A_\theta^G \equiv \tilde{U} B_3 \frac{\partial U}{\partial \theta} = B_2 e^{B_3 \varphi}$$

From another derivative:

$$\frac{\partial U}{\partial \varphi} = B_1 e^{B_3 \varphi} \sin\theta, \qquad A_\varphi^G \equiv \tilde{U} B_3 \frac{\partial U}{\partial \varphi} = \sin^2\theta - B_1 e^{B_3 \varphi} \sin\theta \cos\theta$$

---

[3] I am also using easily verified $e^{-B_3 \varphi} B_2 = B_2 e^{B_3 \varphi}$

We see that full $G_3^+$ values $A_\theta^G$ and $A_\varphi^G$ differ from usual quantum mechanical case by additional bivector terms. The Berry curvature also has additional bivector term:

$$F_{\theta\varphi}^G = \partial_\theta A_\varphi^G - \partial_\varphi A_\theta^G = \sin 2\theta + B_1 e^{B_3\varphi}(1 - \cos 2\theta)$$

The additional bivector terms may be an indicator of a kind of torsion caused by (6.2) $g$-qubit transformation, though it needs further elaboration.

## 7. Geometric evolution.

With explicitly defined variable "imaginary unit" many things become not just more informative but also much simpler. As an example take the $G_3^+$ variant of geometric evolution derived in usual quantum mechanical approach through holonomy (see, for example, [4]).

Transformation (6.1) can be straightforwardly generalized to:

$$so(\alpha, \beta, S) \to e^{\frac{H(t)}{|H(t)|}(|H(t)|)} so(\alpha, \beta, S) \equiv so(\alpha, \beta, S, t)$$

The "complex" plane defined by unit bivector $\frac{H(t)}{|H(t)|}$ then becomes variable and $|H(t)|$ plays the role of angle in this plane. The evolution equation for the result of transformation is:

$$\frac{d}{dt} so(\alpha, \beta, S, t) = \frac{H(t)}{|H(t)|} |H(t)| H'(t) so(\alpha, \beta, S, t),$$

which is a generalization of Schrödinger equation with variable "imaginary unit" $i(t) = \frac{H(t)}{|H(t)|}$.

Let's make assumption that $U(t)$ in $H(t) = \tilde{U}(t) H_0 U(t)$ just generates rotation, so $|H(t)| = |H_0|$. If the initial value of Hamiltonian also lies on $S^3$ we have:

$$\frac{d}{dt} so(\alpha, \beta, S, t) = H(t) H'(t) so(\alpha, \beta, S, t) = \frac{1}{2} \frac{d}{dt}(H^2(t)) so(\alpha, \beta, S, t)$$

with the solution:

$$so(\alpha, \beta, S, t) = e^{-\frac{1}{2} H_0^2} e^{\frac{1}{2} H^2(t)} so(\alpha, \beta, S, 0)$$

8. Conclusions.

The idea of using variable plane in 3D for playing the role of "complex plane", in combination with $G_3^+$ algebra, resulted in more detailed description of quantum states and observables usually formalized in terms of two-dimensional "complex" vectors and Hermitian matrices. New additional components in Berry parameters may be used to analyze possible anyonic states in the topological computing related structures.